\documentclass[11pt]{article}

\usepackage{amsmath,amssymb}
\usepackage{slashed}
\usepackage{xcolor} 
\usepackage{graphicx}
\usepackage{makecell}
\usepackage{dcolumn} 
\usepackage{bm} 
\usepackage{epsfig}
\usepackage{epstopdf}
\usepackage{grffile}
\usepackage{color}
\usepackage{colordvi}
\usepackage{amsmath,amssymb}
\usepackage{rotating}
\usepackage{lscape}
\usepackage{cite}
\usepackage{float}
\usepackage{hyperref}

\def\Re{{\cal R \mskip-4mu \lower.1ex \hbox{\it e}\,}}
\def\Im{{\cal I \mskip-5mu \lower.1ex \hbox{\it m}\,}}

\def\tev{\,{\ifmmode\mathrm {TeV}\else TeV\fi}}
\def\gev{\,{\ifmmode\mathrm {GeV}\else GeV\fi}}
\def\mev{\,{\ifmmode\mathrm {MeV}\else MeV\fi}}

\begin{document}

\begin{center}

\vspace*{15mm}
\vspace{1cm}
{\Large \bf Probing the nonstandard  top-gluon couplings through $t\bar{t}\gamma\gamma$ production at the LHC}

\vspace{1cm}

{\bf Seyed Mohsen Etesami and Esmat Darvish Roknabadi}

 \vspace*{0.5cm}

{\small  School of Particles and Accelerators, Institute for Research in Fundamental Sciences (IPM) P.O. Box 19395-5531, Tehran, Iran } \\
\vspace*{.2cm}
\end{center}

\vspace*{10mm}

%
%
\begin{abstract}\label{abstract}

In this paper, we investigate the anomalous chromoelectric and
chromomagnetic dipole moments
of the top quark through top pair production in association 
with two photons at the LHC. We first present the strategy to
reconstruct this process assuming a different source for
background processes. Then, we focus on the existing constraints from inclusive
top-pair production from the Tevatron and LHC, adding the new LHC
measurement. Afterwards, we introduce the new cross section ratio $R_{2\gamma/\gamma}=\sigma_{t\bar{t}\gamma\gamma}/\sigma_{t\bar{t}\gamma}$
 and show the usefulness of this ratio in canceling most of the systematic
 uncertainties and
its special functionality to constrain dipole moments. Finally, we use the scalar sum of the transverse momentum of  jets, $H_{T}$,
in order to define a signal-dominated region and obtain
limits on these anomalous top couplings using different amounts
of expected data from the LHC.
\end{abstract}

\vspace*{3mm}

PACS Numbers:


\newpage


\section{Introduction}\label{Introduction}
The top quark is to date the heaviest observed elementary particle~\cite{ATLAS:2014wva}. Therefore, from the theoretical point of view
it plays an important role in the electroweak symmetry breaking (EWSB)
mechanism as it has the largest Yukawa coupling among all of
the fundamental particles. Furthermore, the top sector is considered one of the most likely
places that new physics can be probed. There are several models
that predict the existence of new particles that are expected to
perferentially couple
to the top quark. Another
attractive aspect of the top quark is the CP properties of its
interactions with the standard model (SM) fields. CP violation
has a tiny contribution in the SM model through the complex phase of
the Cabibbo-Kobayashi-Maskawa matrix which is not big enough to explain the observed matter-antimatter asymmetry in the Universe and needs a new source of CP
violation which should come from beyond SM. The CP-violating terms in the
top-quark interactions from BSM physics can appear as electric dipole moment (EDM),
chromo-EDM (CEDM), and weak-EDM  terms. Therefore, the precise measurement
of these moments will pave the way for finding the effects of new
physics.

The first and second runs of the LHC with the center-of-mass
energies of 7, 8, and 13 TeV confirmed the SM model of particle physics by
discovering the long sought-after Higgs
boson~\cite{Aad:2012tfa,Chatrchyan:2012xdj}, but and no hint of 
BSM physics has been found. However, there are many SM properties that
have not been measured
accurately yet. Therefore, one of the missions of the phase II upgrade
of the LHC
is to make these measurements precise by providing an unprecedented
amount of data, which is ultimately expected to be 3 $ab^{-1}$ of integrated
luminosity (IL). In this context, many rare SM processes
become accessible, such as multiboson processes (like VVV, VVVV) and the associate productions of top-quark pairs with multibosons (like $t\bar{t}$+VV processes,
where, V stands for W, Z, and $\gamma$). These processes with multiple
fermionic and bosonic degrees of freedom provide a rich ground for
testing the fermion-gauge boson as well as triple and quartic
gauge boson couplings predicted within the SM, also BSM~\cite{Maltoni:2015ena,Etesami:2017ufk}. Even
though the production phase space of these processes is limited due
to the higher
energy threshold (which leads to the lower cross section), they  benefit
from multiplications of final-state particles which significantly reduce 
background contributions.
It should be mentioned that the cross sections of $t\bar{t}W$, $t\bar{t}Z$,
and $t\bar{t}\gamma$ processes have been measured by the CMS and
ATLAS collaborations~\cite{Aaboud:2016xve, Sirunyan:2017uzs,
  Aad:2015uwa, Sirunyan:2017iyh, Aaboud:2018hip}. 

The aim of this paper is to study how well the chromoelectric and  magnetic
dipole moments of the top quark can be measured during top quark pair production in
association with two photons, $t\bar{t}\gamma\gamma$. As these dipole moments are absent at tree level and they
can only show up in higher-order corrections, they turn out to be
very small in the SM. Therefore, any deviation could indicate the
presence of new physics, on the contrary, consistency with the SM values could
constrain the new couplings that may contribute to this
process. 

The paper is organized as follows. In Sec.~\ref{effL}, we describe
an effective field theory approach and define the top-quark CEDM and
chormomagnetic dipole moment (CMDM)
in this context, and we translate of these moments to dimension-six operators. In 
Sec~\ref{sm}, we explain  $t\bar{t}\gamma\gamma$ production at the LHC within
the SM framework, and then use the dimension-six operators
via the effective Lagrangian approach to calculate the cross
section. In Sec~\ref{analysis}, we explain the signal process
selection strategy and consider related
background processes. In Sec~\ref{constsec}, we discuss the current
constraints on $d_{V}^{g}$ and $d_{A}^{g}$ from inclusive top pair
production; then, we introduce  the new ratio 
$R_{2\gamma/\gamma}=\sigma_{t\bar{t}\gamma\gamma}/\sigma_{t\bar{t}\gamma}$
to constrain the anomalous couplings. In Sec~\ref{htsec}, we
employe the scalar sum of the jet's transverse momentum distribution to
define a signal-dominated region and use a single-bin experiment
to extract the limits on $d_{V}^{g}$ and $d_{A}^{g}$
respectively. Finally, in Sec~\ref{summary},
we summarize the results and conclude the paper.

\section{$gt\bar{t}$ effective coupling}\label{effL}

Effective field theory is a remarkable framework to describe the
effects of physics at a high energy scale $\Lambda$, which is
necessarily higher than the energy scale of the experiment. Essentially when the heavy degrees of freedom from high-energy physics
can not be directly produced one can 
integrate them out, resulting in new terms which are added to the SM
Lagrangian. These new terms are composed of higher-dimension operators
suppressed by the inverse power of $\Lambda$, and
they respect Lorentz invariance, SM gauge
symmetries, and baryon, and lepton number conservations. Thus, the SM effective
Lagrangian up to the dimension-six operators can be written as follow:

\begin{eqnarray}
\mathcal{L}_{eff} = \mathcal{L}_{\rm SM} + \sum_{i}\frac{c_{i}\mathcal{O}^{(6)}_{i}}{\Lambda^{2}},
\end{eqnarray}

where $\mathcal{L}_{\rm SM}$ is the SM
Lagrangian. $\mathcal{O}^{(6)}_{i}$  are the dimension-six
operators (which are the dominant contribution to
the experimental observables) and the $c_{i}$'s are unknown dimensionless coefficients
that describe the strength of the new physics couplings to the
SM particles. After EWSB, the integrated-out terms will produce new
couplings that do not exist at tree level in the SM (such as
electric and magnetic dipole moments), as well as
couplings which correct the SM interactions. The most general form of
the g$t\bar{t}$ coupling assuming up to dimensions-six operators
can be depicted as follows:

\begin{eqnarray}\label{e1}
\mathcal{L}_{gt\bar{t}} = -g_{s}\bar{t}\frac{\lambda^{a}}{2}\gamma^{\mu}tG^{a}_{\mu}-g_{s}\bar{t}\frac{\lambda^{a}}{2}\frac{i\sigma^{\mu\nu}}{m_{t}}
(d_{V}^{g}+id_{A}^{g}\gamma_{5})tG^{a}_{\mu\nu},
\end{eqnarray}

ًwhere $g_{s}$, $\lambda^{a}$, and $G^{a}_{\mu\nu}$ are the strong
coupling constant, Gell-Mann matrices, and gluon field-strength
tensor, respectively. $d_{V}^{g}$ and $d_{A}^{g}$ are real parameters
which represent the top-quark chromomagnetic and chromoelectric dipole
moments. The first term is the SM interaction, while the
rest of the terms contain the $gt\bar{t}$
and $ggt\bar{t}$ interactions and are generated from dimension-six
operators based on the convention used
in~\cite{Buchmuller:1985jz,AguilarSaavedra:2008zc}, which have the
following form:

\begin{eqnarray}\label{e2}
 O^{33}_{uG\phi} \sim (\bar{q}_{L3}\lambda_{a}\sigma^{\mu\nu}t_{R})\tilde{\phi}G^{a}_{\mu\nu},
 \end{eqnarray}

where $\bar{q}_{L3}$ and $t_{R}$ are the weak doublet of the left-handed
quark field and right-handed top quark field respectively. $\phi$
is the weak doublet of the Higgs field and $\tilde{\phi}=
i\tau_{2}\phi^{*}$.
The relation between the dimension-six operator in equation~\ref{e2}
and the chromo-moments of the top quark after the symmetry breaking can be written
as: 

\begin{eqnarray}\label{cgtt}
\delta d^{g}_{V} = \frac{\sqrt{2}}{g_{s}} {\rm Re}O^{33}_{uG\phi}\frac{vm_{t}}{\Lambda^{2}}~,~\delta d^{g}_{A} = \frac{\sqrt{2}}{g_{s}}{\rm Im}O^{33}_{uG\phi}\frac{vm_{t}}{\Lambda^{2}},
\end{eqnarray}

where $m_{t}$ is the top-quark mass and $v$ is the vacuum expectation value
of the Higgs field. The CEDM and CMDM are related to the real and
imaginary part of $O^{33}_{uG\phi}$ and both are considered in this
study.

In the SM, the CMDM of the top quark ($d_{V}^{g}$) can be generated via one-loop
QCD and electroweak diagrams. There are two types of Feynman diagrams that
contribute to the QCD part. The first diagram is the one with an external
gluon emitted from the internal top quark and in the second diagram
the external gluon is comes from the exchanged gluon due to the non-Abelian
properties of the strong interaction. The total QCD contribution
is $d_{V}^{g}=-0.008$~\cite{Martinez:2007qf}, which is the dominate SM loop contribution.
In the electroweak loop diagrams $W^{\pm}$, $Z$, and Higgs bosons can be exchanged
while the gluon coming from the internal quark. This tiny contribution is
about 12$\%$ of the QCD part but with the opposite sign. Finally, the total SM loop correction is
$d_{V}^{g}=-0.007$~\cite{Martinez:2007qf}. The CEDM contribution in the SM
arises from the three-loop diagrams and has been shown to be very
small~\cite{Martinez:2007qf}.

Direct bounds on the CMDM and CEDM from inclusive
and differential measurements of  $t\bar{t}$ processes at the Tevatron
and LHC have been obtained~\cite{Hioki:2009hm, Hioki:2013hva,
  HIOKI:2011xx, Kamenik:2011dk, Hesari:2012au,
  AguilarSaavedra:2018nen, Franzosi:2015osa, Bylund:2016phk}. Also, with the
considerable amount of the data that the LHC in its upgraded phase
will collect, the rare SM processes such as $t\bar{t}$ in association with
two heavy gauge bosons and multi top quark production have been shown to be sensitive to these anomalous
interactions of the top quark and gluon~\cite{Etesami:2017ufk,
  Malekhosseini:2018fgp}. In addition, the CMS
experiment has obtained the limits on these dipole moments via the
measurement of $t\bar{t}$ spin correlation
using the $\sqrt{s}=8$ TeV data~\cite{CMS:2014bea}. Moreover, it has
been shown that there is sufficient sensitivity to probe the CMDM and
CEDM given the high invariant mass of top pair processes where top quarks are highly boosted~\cite{Aguilar-Saavedra:2014iga}. 
Single top tW production has also been shown to be sensitive to top quark chromomoments via its cross section and top-quark
polarization~\cite{Ayazi:2013cba,Rindani:2015vya}.

In addition to the direct bounds, there are also indirect bounds on the top-quark dipole
moments from low-energy measurements which are known to be the
stringent limits up to now. For example, from the measurement of rare $B$-meson
decays  $b\rightarrow s\gamma$, the top-quark chromomagnetic
moment constrains at 95$\%$ confidence level (CL) with $-3.8 \times 10^{-3} < d_{V}^{g}<
1.2 \times 10^{-3}$ ~\cite{Martinez:2001qs}. Also, measurement of the neutron electric
dipole moment could constrain the top-quark chromoelectric dipole
moment to $|d_{A}^{g}|\le 0.95 \times 10^{-3}$ at 90$\%$ 
CL~\cite{Kamenik:2011dk}. In the next sections, we examine the
potential of the $t\bar{t}\gamma\gamma$ process to probe the top-quark
CMDM and CEDM.

\section{$t\bar{t}\gamma\gamma$ production in proton-proton collisions}\label{sm}

\begin{figure}[htb]
\begin{center}
\vspace{1cm}
\resizebox{0.3\textwidth}{!}{\includegraphics{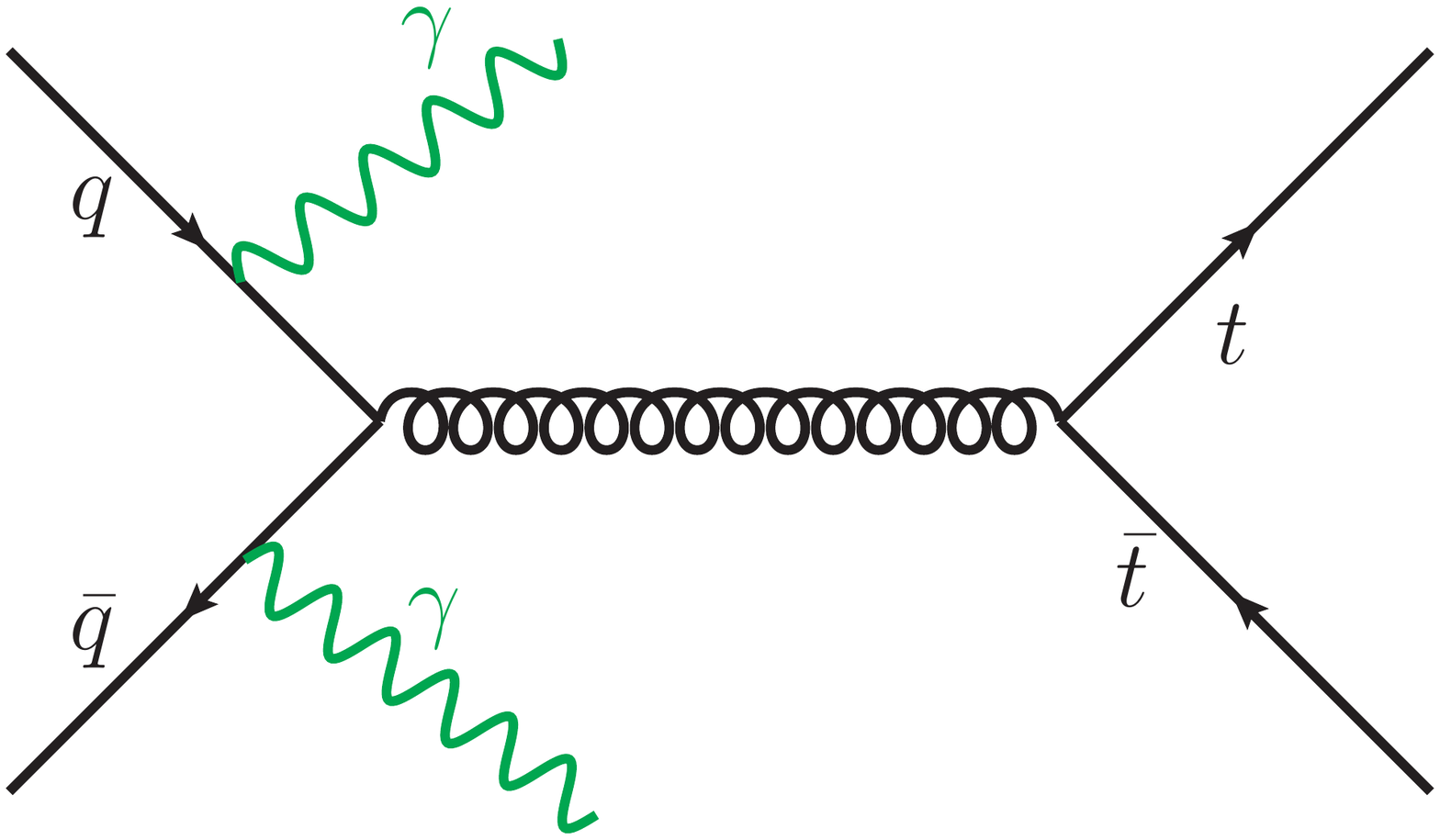}}  
\resizebox{0.3\textwidth}{!}{\includegraphics{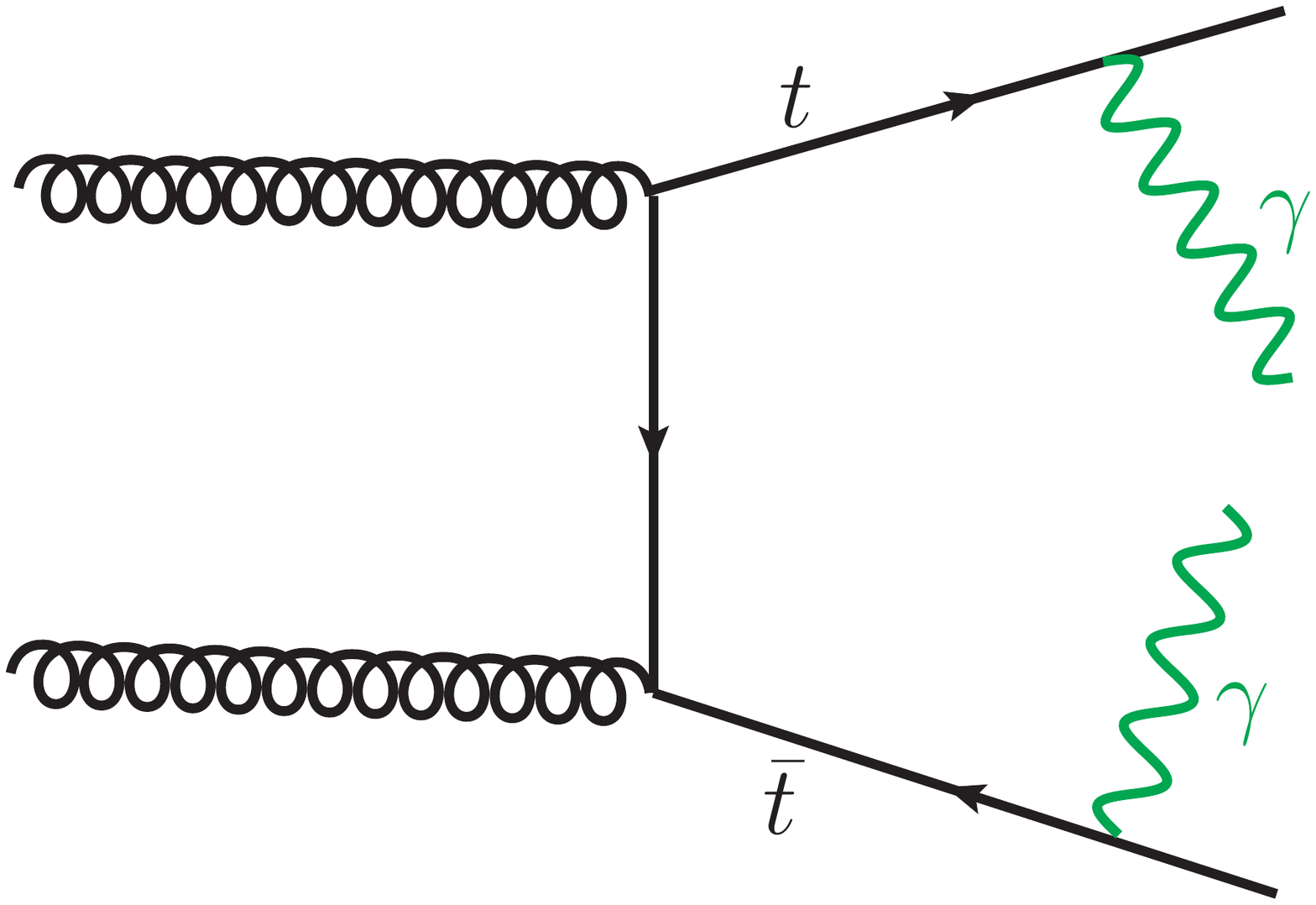}}  
\resizebox{0.3\textwidth}{!}{\includegraphics{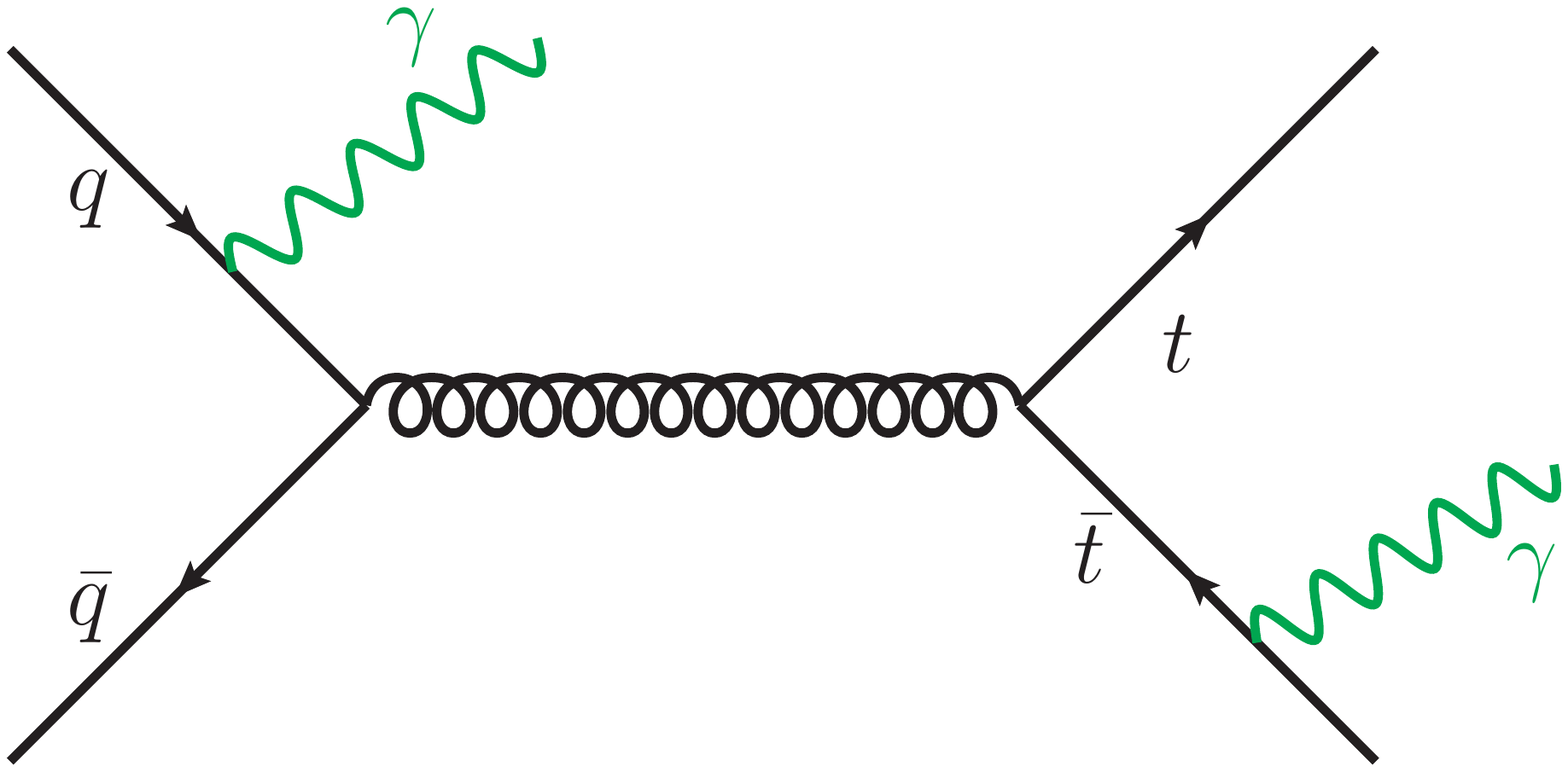}}  

\caption{  Dominant Feynman diagrams for $t\bar{t}$ production  in
  association with two photons within the SM framework.}\label{diagrams}
\end{center}
\end{figure}

Top-pair production in association with two photons within the SM
framework can occur through gluon-gluon fusion or quark-antiquark annihilation at the LHC. The Feynman diagrams with the dominant
contribution for the 
$t\bar{t}\gamma\gamma$ process are shown in 
Figure~\ref{diagrams}. The reason that the dominant production mode for $t\bar{t}\gamma\gamma$
is from $q\bar{q}$ annihilation comes from the fact that photons can radiate either from top quarks or initial-state quarks, while this is not
the case for the gluon-gluon fusion production mode. For instance, the calculated 
contributions of the  $q\bar{q}$ production mode at leading order (LO) with $\sqrt{s}=13$ TeV for 
$t\bar{t}$, $t\bar{t}\gamma$, and $t\bar{t}\gamma\gamma$ processes are
13$\%$, 32$\%$, and 66$\%$, respectively when the $p_{T}$ of the
photon is set greater than 10 GeV at the generator level.

We use the \texttt{MadGraph5\_aMC@NLO} package \cite{mg5} for event
generation and to calculate the cross sections of signal and relevant
backgrounds. The total cross section is calculated assuming the top-quark
mass $m_{t}$= 172.5 GeV, W-boson mass $m_{W}$= 80.37 GeV and
$G_{F}=1.16639\times10^{-5}  {GeV^{-2}}$. The NNPDF3 PDF is used as 
parton distribution functions~\cite{nnpdf}. The values of the factorization
scale ($\mu_{F}$) and renormalization scale ($\mu_{R}$) are calculated
event by event and are considered to be
$\mu_{F}=\mu_{R}=\sqrt{m_{t}^2+\Sigma_{i}p_{T}^2(i)}$,  where the sum
is over the visible final-state particles. Top-quark and W-boson decays
are considered in the narrow-width approximation and spin correlation
in the top quarks decay is considered.

To calculate the cross section of $t\bar{t}$+X  including the
chromomoments of the top
quark, the effective Lagrangian is imported via the \texttt{FeynRules}
package~\cite{feyn} and the obtained UFO model~\cite{ufo} is linked to \texttt{MadGraph5\_aMC@NLO}
in order to generate events and calculate the cross
section. The total calculated cross section at leading order arising
from dimension-six operators in the equation~\ref{e2}, including
the CMDM and CEDM of the top quark, is
parametrized as

\begin{eqnarray}\label{sigfunc}
\sigma_{total}=\sigma_{SM}+\alpha~d_{V}^{g}+\beta~(d_{V}^{g})^{2} + \kappa~(d_{A}^{g})^2,
\end{eqnarray}

where $\sigma_{SM}$ is the SM cross section. The second term is
the interference between the SM and the real part of the $O^{33}_{uG\phi}$
operator, which has a $\frac{1}{\Lambda^{2}}$ contribution. The
third and forth quadratic terms correspond to the pure real and
imaginary parts of $O^{33}_{uG\phi}$ which have the contributions of
the order of $\frac{1}{\Lambda^{4}}$. It should be
mentioned that dimension-eight operators also generate additional
terms of  the
order of $\frac{1}{\Lambda^{4}}$, but we drop those terms as we only consider the dimension-six operators in this analysis. In the equation~\ref{sigfunc}, there is no linear $d_{A}^{g}$ term as the
cross section must be a CP-conserving  observable.

In addition to the signal process, we generate several
reducible background processes using ~\texttt{MadGraph5\_aMC@NLO}, such as $t\bar{t}\gamma$,
$W\gamma\gamma$+jets, single top+$\gamma\gamma$, Z$\gamma\gamma$+jets, diboson+$\gamma\gamma$
including WW$\gamma\gamma$, WZ$\gamma\gamma$, and ZZ$\gamma\gamma$
and finally $\gamma\gamma$+jets, as well as an irreducible background
which is the SM $t\bar{t}\gamma\gamma$. All of the generated samples are passed to \texttt{Pythia 8} \cite{pythia8}
in order to perform parton showering and
hadronization. Jet clustering is performed using the anti-$k_{t}$
algorithm~\cite{Cacciari:2008gp} implemented in the \texttt{FastJet}
package~\cite{Cacciari:2011ma} using a radius parameter of R=0.5. b-tagging
and mistagging efficiencies for the jets that originate from
the hadronization of a b-quark are considered\cite{Sirunyan:2017ezt}. These efficiencies are
parametrized based on the transverse momentum of the jets. In this
analysis,  the fast detector response is estimated
using the \texttt{Delphes 3.4.1} package~\cite{deFavereau:2013fsa}
based on the similar conditions of the CMS detector.
\section{Analysis strategy}\label{analysis}

In this section, we present the analysis strategy to select the
$t\bar{t}\gamma\gamma$ signal events. We also discuss the relevant background processes and
estimate their contributions in this final state. As a result, one
can obtain the potential power of this process to probe the
chromomoments of top quark, which we will discuss in the next
sections. In this analysis we consider the semi-leptonic decay
mode of the $t\bar{t}\gamma\gamma$ process, as this decay mode has the large
contribution and the presence of one lepton along with two
photons will help to effectively suppress the background processes.

In order to select signal events we require to have exactly 
two isolated photons with transverse momenta $p_{T}>$ 25 GeV and
pseudorapidity of $|\eta|<$ 2.5. We also demand to have a
lepton (electron or muon) with the same $p_{T}$ and $\eta$ cut
values as the photons. Moreover, we veto events that contain
any other leptons in order to suppress the backgrounds, including
Z-boson events such as Z$\gamma\gamma$+jets. The requirements for jet
selection are  $p_{T}>$ 40 GeV and $|\eta|<$ 5, and the requirements for the b-tagged jets
are $p_{T}>$ 40 GeV and $|\eta|<$ 2.5. In order to suppress the backgrounds
that do not contain a W boson, we require the missing transvers
energy (MET) to be greater than 30 GeV. In addition to the mentioned cuts, in
order to have well-isolated objects we require the angular separation
between the two photons and between the photons and other
objects to be  $\Delta R(\gamma, X)=\sqrt{\Delta \phi^2+\Delta \eta^2}>$ 0.5 where
X=e, $\mu$, jets, b-jets, or $\gamma$. On top of the other requirements, we
ask the events to have $H_{T}>$ 300 GeV
where this variable is defined as $H_{T}=\Sigma~p_{T}$, and the sum is
performed over the transverse momentum of jets within the defined
acceptance region. Higher values of $H_{T} $ correspond
to the heavier final state masses which is $t\bar{t}$ in the case of
our signal and could suppress the processes with lower-mass and no-mass states such
as W+jets+$\gamma\gamma$ and $\gamma\gamma$+jets respectively. Table~\ref{cutflow} shows the
expected yields for the two signal samples $d_{V}^{g}$=0.2,
and $d_{A}^{g}$=0.2 as well as the SM backgrounds after applying each
set of selection cuts. It should be mentioned that the expected yields for
the signal samples comprise the contribution of anomalous top-quark
dipole moments, the SM $t\bar{t}\gamma\gamma$ contribution, and their
interference. 
\begin{table}[]
\centering 
\caption{Expected number of events for the two signal samples and
  backgrounds after applying the selection cuts for 100 $fb^{-1}$ IL.}
\label{cutflow}
\begin{tabular}{c|cc|c c c c c c} \hline\hline
\small{$\sqrt s = 13$ TeV, 100$fb^{-1} IL$}      &
                                           \multicolumn{2}{c|}{~\small{Signal+$t\bar{t}\gamma\gamma$}
                                           }    &
                                                  \multicolumn{3}{c}{\,\,\,\,\,~~~~
                                                  \small{Backgrounds
                                                  with 2 real photons} }    \\   \hline

\small{Selection cuts} & \small{$d_{V}^{g}$=0.2} & \small{$d_{A}^{g}$=0.2} & \small{$t\bar{t}\gamma\gamma$}& \small{$W/Z+jets\gamma\gamma$}&\small{ST$\gamma\gamma$}&\small{diboson$\gamma\gamma$}& \small{jets$\gamma\gamma$}\\ \hline\hline
\small{\makecell{$p_{T,lep},~\eta_{lep}$~$N_{lep}=1$\\, lepton$\&\gamma$ veto}}&\small{305}&\small{582}&\small{206}    &\small{1834}  &\small{239}& \small{6491} & \small{1.36$\times 10^{5}$} \\ \hline
\small{$p_{T,\gamma},~\eta_{\gamma}$,~$N_{\gamma}=2$} &\small{47}&\small{87} &\small{29}  &\small{147}&\small{39}&\small{50}&\small{281}\\ \hline
\small{\makecell{$p_{T, jets, b-jets},~\eta_{jets,b-jets}$,\\
  $N_{jets}>1$,$N_{b-jets}=2$, MET\\, $H_{T}$, $\Delta R(\gamma, X)$} } &\small{7.15}& \small{10.20} & \small{3.01}   & \small{0.03}  &\small{0.44}& \small{0.09}& \small{0}  \\ \hline
\end{tabular}
\end{table}

Apart from the background processes with two real
photons, there is a contribution from the $t\bar{t}\gamma$
process. First, it should be mentioned that $t\bar{t}\gamma$ is a
different process from $t\bar{t}\gamma\gamma$ as the latter has two
photons in the matrix elements while the former has one. However,
events from the $t\bar{t}\gamma$ process may have overlap with
$t\bar{t}\gamma\gamma$ when the showered photon by \texttt{Pythia}
lands into the generator acceptance of $t\bar{t}\gamma\gamma$. Due to
the high cut value applied for $\Delta R$ between the photons and other
objects, one would expect this overlap to be small. However, we have
subtracted this contribution in order to be precise in our
background estimation. The total obtained yield for  $t\bar{t}\gamma$
after applying all of the selection cuts is 14 for 100 $fb^{-1}$.

In addition to the above background processes, in the
real experiment there is a probability that the jets are misidentified 
as a photon. The reason behind this misidentification is that inside
a jet there is a considerable amount of neutral hadrons
such as pions which promptly decay into the two photons in the boosted
topology. Therefore, the produced shower of these two close-by photons
will overlap inside the electromagnetic calorimeter and be misidentified
as a photon, a so-called fake photon. As a result, in a real detector,
processes with large cross sections, such as W/Z+jets, W/Z+jets+$\gamma$,
multijet+$\gamma$, and multijet, may pass our selection
criteria due to this mis-reconstruction of jets. In real experiments such as CMS and ATLAS the probability of
jet-to-photon mis-reconstruction, $P_{j\rightarrow\gamma} $  varies between $10^{-3}-10^{-5}$ depending on the transverse momentum and
pseudorapidity of a photon. We have estimated the contribution of these processes by applying the
selection cuts explained in the previous section except the photon.
Then the resulting cross sections are multiplied by $P_{j\rightarrow\gamma}$ or  $P_{j\rightarrow\gamma}^2 $  according to the number of
mis-reconstructed photons for each process. The contribution of these processes is found to be negligible. However, precise estimation of fake photons is usually performed
using data-driven techniques and a full simulation of detector
components which is beyond the scope of this analysis.

\section{ $t\bar{t}\gamma\gamma$ process's role in constraining the strong dipole
  moments of the top quark}\label{constsec}
In this section we explain how the $t\bar{t}\gamma\gamma$ process can
play a complementary role in constraining the chromo-moments of top
quark. First we discuss the current bounds that one can
obtain from the inclusive cross section measurements of $t\bar{t}$+X
where X=$\gamma$, jets. Then, in the second part we discuss the
constraints from the newly defined ratio 
$\sigma_{t\bar{t}\gamma\gamma}/\sigma_{t\bar{t}\gamma}$  and the
possible improvements with respect to inclusive $t\bar{t}$ cross section
measurements. 

\subsection{Constraints from $t\bar{t}$+X production measurements}

In order to obtain stringent bounds on $d_{V}^{g}$ and $d_{A}^{g}$
one could combine results from different
experiments. In this section we consider the measurements on the
inclusive cross sections of $t\bar{t}$ at the Tevatron from $p\bar{p}$
collisions at $\sqrt{s}$=1.96 TeV~\cite{Aaltonen:2013wca}, the combined
measurements of pp collisions at CMS and ATLAS with $\sqrt{s}$=8 TeV~\cite{ATLAS:2014aaa}, and two other recent
measurements at CMS on the cross section of $t\bar{t}\gamma$ at
$\sqrt{s}$=8 TeV~\cite{Sirunyan:2017iyh} and the cross section of
$t\bar{t}$ at $\sqrt{s}$=13 TeV~\cite{Sirunyan:2017uhy}.

\begin{figure}[htb]
\begin{center}
\vspace{1cm}
\resizebox{0.45\textwidth}{!}{\includegraphics{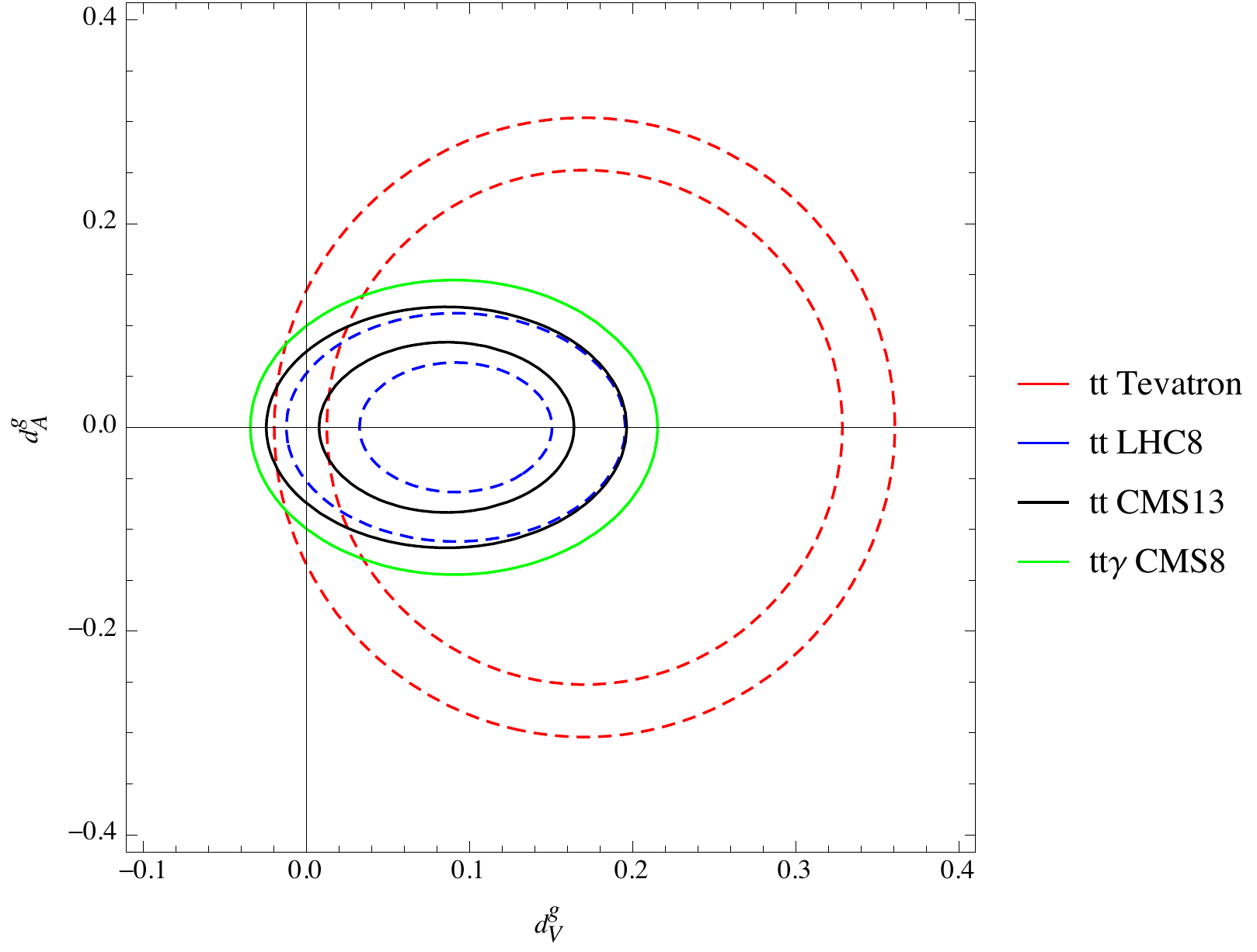}}  
\resizebox{0.50\textwidth}{!}{\includegraphics{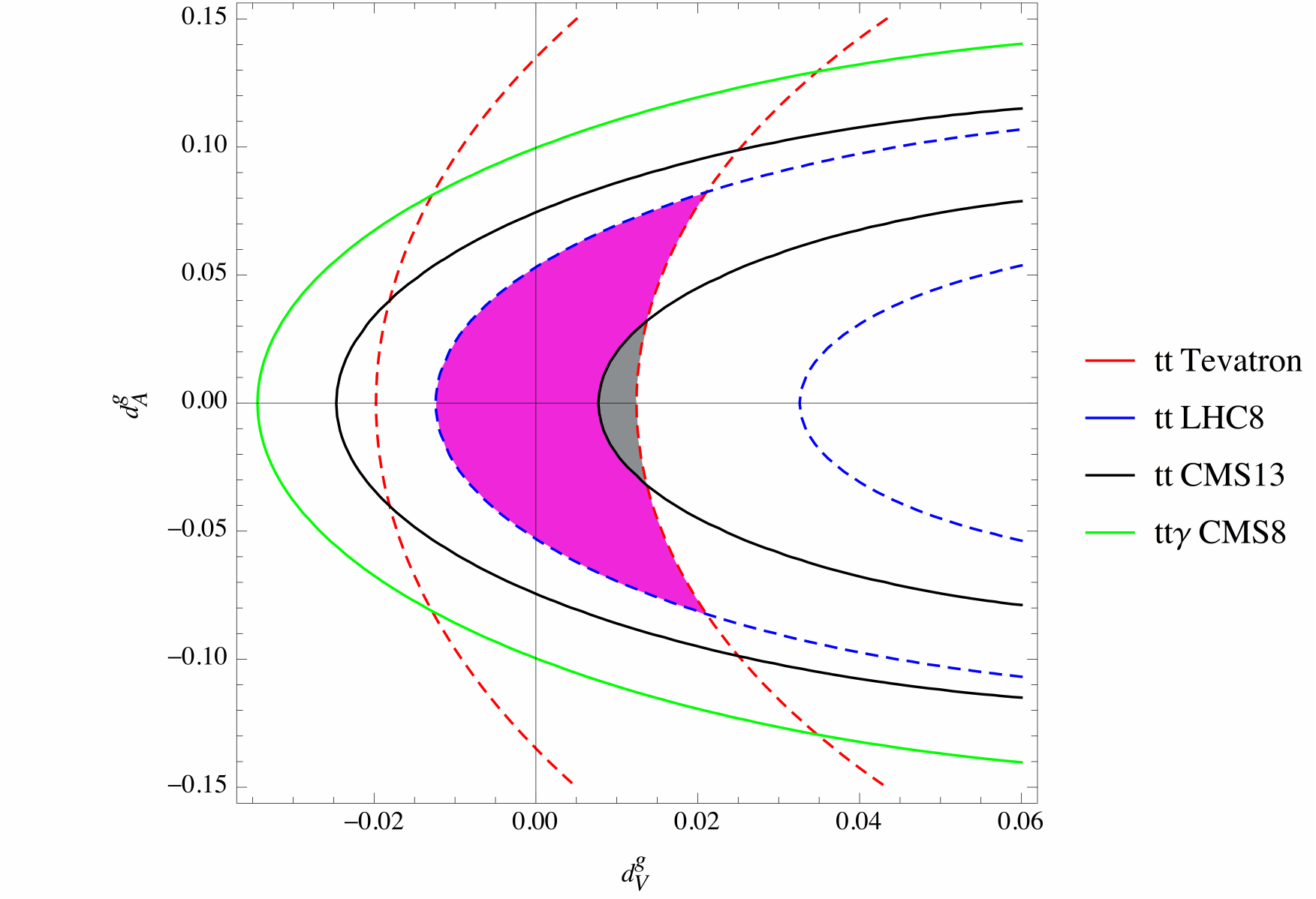}}  
\caption{  Two-dimensional allowed regions for $d_{V}^{g}$ and
  $d_{A}^{g}$  using the different measurements that have been
  performed so far. The left panel shows each experimental bound and the right panel
  depicts the overlap region in the zoomed view.}\label{ttbarcont}
\end{center}
\end{figure}

In order to obtain the experimental bounds, one needs to drive the
functionality of the total cross section to the anomalous couplings
based on the particles that collide and the center-of-mass energy of collisions. Exploiting the method explained in 
section~\ref{sm}, we evaluate the total cross sections including the
leading-order contribution of top quark
chromo-moments. Table~\ref{ttcof} shows the obtained values for each coefficient
belonging to the linear and quadratic terms for each measurement. Then,
the constraints obtained using the measured values for the total cross
sections along with the precise available cross sections that the SM predicts at
next-to-leading order (NLO) or next-to-next-to-leading order (NNLO)
calculated with
\texttt{Top++}~\cite{Czakon:2011xx}. In order to obtain the limit
bands we consider the total uncertainty of cross sections measured
in ~\cite{Aaltonen:2013wca, ATLAS:2014aaa, Sirunyan:2017iyh, Sirunyan:2017uhy}, as well as the theoretical uncertainties on the predicted cross sections including
the PDF, renormalization/factorization scales, and top-quark mass
uncertainties. The theoretical uncertainties for the $t\bar{t}$ cross
section at NNLO (calculated in Ref. \cite{Czakon:2011xx})  arising from
scale variations and (PDF$+\alpha_{s}$) are 3 and 5$\%$ for  $\sqrt{s}$=8 TeV, 3 and
4$\%$ for  $\sqrt{s}$=13 TeV, and 2 and 
2$\%$ for $\sqrt{s}$=1.96 TeV,
respectively.  The left panel of Figure~\ref{ttbarcont} depicts the two-dimensional bounds on $d_{V}^{g}$ and $d_{A}^{g}$ for each
measurement separately, and the right panel shows the overlap region
of all measurements in zoomed view. The total colorful area is the bound obtained at Tevatron and
LHC8, which is compatible with the results of Refs~\cite{Hioki:2013hva,
  Aguilar-Saavedra:2014iga}. The pink area is the bound obtained
by adding CMS13 $t\bar{t}$ cross section measurements with 2.2 $fb^{-1}$
integrated luminosity and the $t\bar{t}+\gamma$ measurement at CMS at
 a center-of-mass energy of 8 TeV. As can be seen
the bounds' improvement is not significant and is only for
the $d_{V}^{g}$ coupling. The $t\bar{t}+\gamma$ measurement also does
not produce tighter bounds. As the $t\bar{t}$ inclusive cross section measurement at CMS13 is
systematically dominated, adding more data recorded by CMS in 2016 will
not increase the precision of the measurements by a large value, and
consequently no big improvement in the bounds of the top-quark dipole moments
is expected. Moreover, one can show that considering better precision of the inclusive
cross section only leads to obtaining better bounds on
$d_{V}^{g}$. Therefore, we introduce a new observable in the selected
phase space which provides a different functionality and can
be used to tighten the current bounds, especially on  CP-violating coupling.

\begin{table}[]
\centering
\caption{Values of $\alpha$, $\beta$, and $\kappa$  for the Tevatron and
  LHC. The SM cross sections are inclusive ones except for the
  $t\bar{t}\gamma$ process which is presented
  per semileptonic final state.}
\label{ttcof}
\begin{tabular}{lllll} \hline\hline
$Process$ & $\sigma_{SM}[pb]$&$\alpha$ & $\beta$ & $\kappa$ \\ \hline\hline
$t\bar{t}$ Tevatron $\sqrt{s}=1.96$ TeV  & $7.35\pm0.21$  &     $-55.9$     &     164       &  64  \\
$t\bar{t}$ LHC $\sqrt{s}=8$ TeV       &  $252.8\pm14.4$  &  $ -1668$       &     9013     &  7828 \\ 
$t\bar{t}$ LHC $\sqrt{s}=13$ TeV     &  $832\pm43$  &  $-5395$     &     31387   &  27400\\
$t\bar{t}\gamma$ LHC$\sqrt{s}=8$ TeV &$0.592\pm0.077$&$-3.39$ &18.75 &13.98\\
\end{tabular}
\end{table}

\subsection{Cross section ratio}

As discussed in the previous section the current cross section
measurements of top-quark pair production have a limited ability to tighten the
current bounds on top-quark chromo-moments. Therefore, we propose a new observable which is the cross section ratio
of $t\bar{t}\gamma\gamma$ to $t\bar{t}\gamma$ within a selected
phase space in order to constrain the currently allowed region
of these anomalous couplings. It is defined as

\begin{eqnarray}\label{ratio}
    R_{2\gamma/\gamma}=\sigma_{t\bar{t}\gamma\gamma}/\sigma_{t\bar{t}\gamma},
\end{eqnarray}

where the numerator and denominator are calculated in the same signal
region except for the required number of photons, which is two for
the numerator and one for the denominator. Using this ratio has the advantage of canceling the
systematic uncertainties. From the theoretical point of view,
uncertainties from the parton distribution function and  $\alpha_{s}$ can be reduced
assuming the leading-order or higher-order corrections. Apart from
that, several systematic uncertainties arising from the luminosity, jet energy scale,
b-jet tagging, and lepton identification can be canceled out. In
particular, using the proposed $R_{2\gamma/\gamma}$ will effectively reduce the photon identification
uncertainties as well.  

There are several studies which are shown the idea of using the cross section
ratio in order to reduce the uncertainties. For instance, in the
Ref~\cite{Plehn:2015cta} the authors showed that by using
the ratio $\sigma_{t\bar{t}+H}/\sigma_{t\bar{t}+Z}$ the top Yukawa
coupling can be measured with 1$\%$ precision using proton-proton collision data with the center-of-mass energy of 100 TeV. In
another study it was shown that at the LHC the cross section ratio of
single top quark production in association with a photon over single top quark production $\sigma_{tj\gamma}/\sigma_{tj}$ is a
precise observable that can probe the top-quark electric and magnetic
dipole moments~\cite{Etesami:2016rwu}. Also, it has been shown that
using $\sigma_{t\bar{t}+\gamma} / \sigma_{t\bar{t}}$ and
  $\sigma_{t\bar{t}+Z }/ \sigma_{t\bar{t}}$ could cancel several sources of
    uncertainties, and these ratios may be more sensitive observables of
  the electroweak dipole operators of the top
  quark~\cite{Schulze:2016qas}. The available measurements on the cross section ratio 
  $\sigma_{t\bar{t}+\gamma} / \sigma_{t\bar{t}}$ at the Tevatron and LHC
    and the measured cross section of a single top quark in association
    with a photon can be found in Refs.~\cite{Aaltonen:2011sp, Aad:2015uwa,Sirunyan:2017iyh, Sirunyan:2018bsr}.

We test $R_{2\gamma/\gamma}$ against the
variation of renormalization and factorization scales by generating dedicated samples considering 
$\mu_{f}=\mu_{R}$ and equate them first to $2\times Q_{0}$
and then to $Q_{0}/2$. Then the ratios for each value of $\mu_{f}=\mu_{R}$
are calculated respectively. The uncertainty
due to this scale variation of $R_{2\gamma/\gamma}$ is obtained
below $\pm0.5\%$ while the uncertainty for
each total cross section is about 12$\%$. We also evaluate the robustness of
these ratios against the variation of parton distribution
functions (PDFs), by generating the different samples for
$t\bar{t}\gamma\gamma$ and $t\bar{t}\gamma$ using the three different
PDF sets NNPDF3.0~\cite{Ball:2014uwa},
MSTW08~\cite{Martin:2009iq}, and CTEQ6L1~\cite{Pumplin:2002vw}. Then,
we calculate the ratio $R_{2\gamma/\gamma}$ for each set of PDFs,
which results in an
uncertainty of about 2$\%$. The stability of this ratio against different
uncertainties shows that this is a robust experimental observable. 

\begin{figure}[htb]
\begin{center}
\vspace{1cm}
\resizebox{0.45\textwidth}{!}{\includegraphics{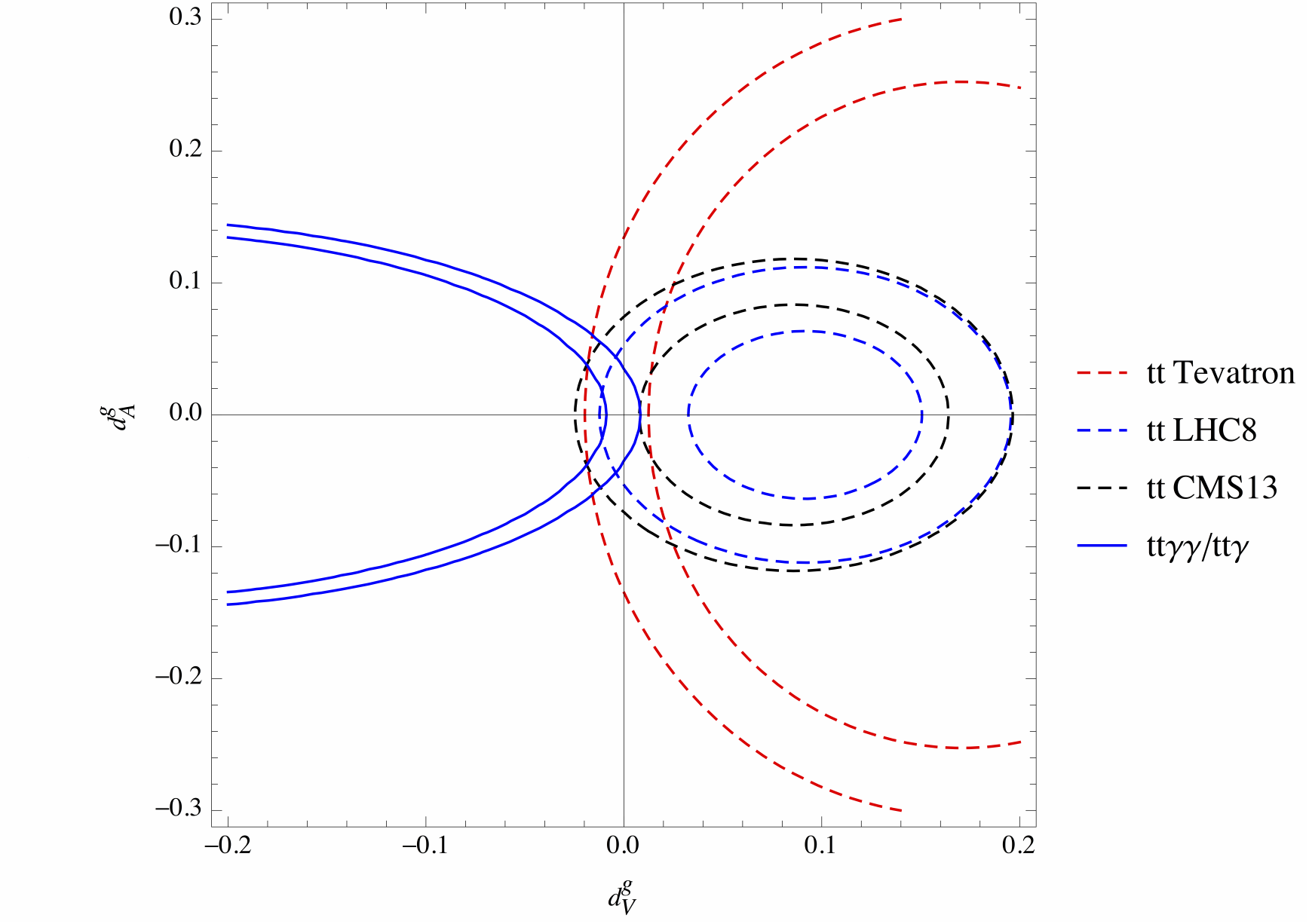}}  
\resizebox{0.54\textwidth}{!}{\includegraphics{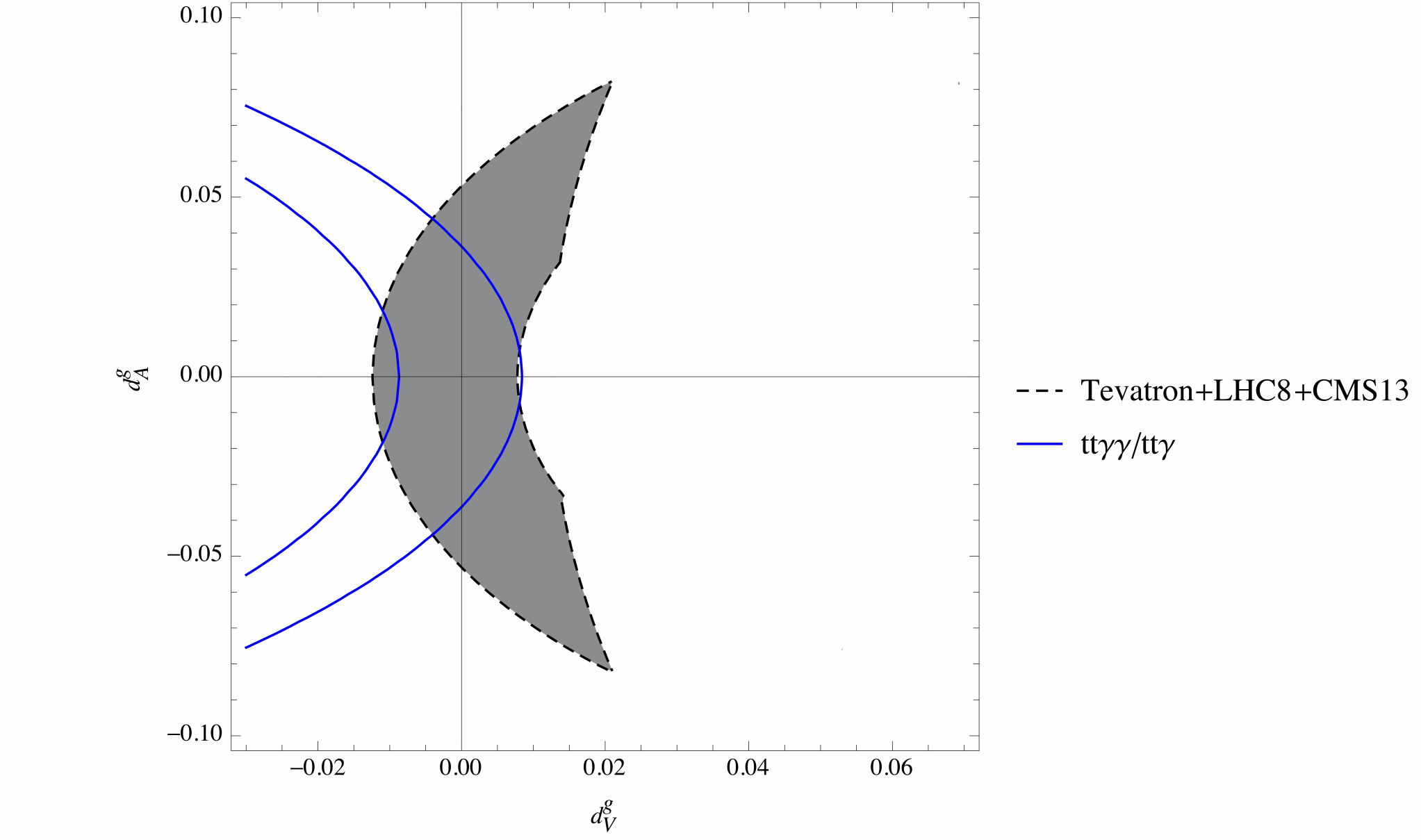}}  

\caption{  Left: Comparison of the current obtained bounds from the Tevatron and LHC at 8
  and 13 TeV using inclusive top-pair production with the extracted
  95$\%$ C.L. allowed region obtained from
$R_{2\gamma/\gamma}$  assuming a 5$\%$ uncertainty in the selected
phase space. Right: A zoomed-in view of the overlap region, showing the improvement in the
obtained constraints
from $R_{2\gamma/\gamma}$ .}\label{crossratios}
\end{center}
\end{figure}

In the following, we discuss the effect of these anomalous couplings on the
defined ratio. As explained in section~\ref{sm}, the
contribution of gluon-gluon fusion to $t\bar{t}\gamma\gamma$ is 
lower than the $t\bar{t}+\gamma$ and $t\bar{t}$ processes, considering
the photon radiation from this initial state is forbidden. Thus, the ratio
$R_{2\gamma/\gamma}$ benefits from this
dissimilar functionality and can probe these anomalous couplings in a region that is different
from the one obtained from the normal inclusive cross
section. Considering the 3 $ab^{-1}$ of integrated
luminosities expected to be delivered by the LHC and the cancellation of
different sources of uncertainty, this ratio in the selected phase
space can be measured with very good precision. Therefore, we
consider the two total uncertainties, 5$\%$  and 10$\%$, and extract the two dimensional 95$\%$ bounds on 
$d_{V}^{g}$ and $d_{A}^{g}$. The left panel of Figure~\ref{crossratios} shows the 95$\%$ C.L. allowed
regions extracted from different measurements (dashed lines) compared
to those obtained from $R_{2\gamma/\gamma}$ with a 5$\%$
uncertainty (solid blue lines).  The right panel of Figure~\ref{crossratios} compares the
current combined limit obtained from the Tevatron and LHC at 8 and 13 TeV (shaded
gray area) with the bound obtained using 
$R_{2\gamma/\gamma}$ with a 5$\%$ uncertainty in the zoomed view. It can be seen that the new
behavior of this ratio can tighten 
the currently allowed region for both anomalous couplings; in
particular, it has a considerable ability to constrain $d_{A}^{g}$. The obtained
bounds using  $R_{2\gamma/\gamma}$  for each coupling are $-0.0088<d_{V}^{g}<0.0083$ and
$-0.037<d_{A}^{g}<0.037$ assuming a 5$\%$ uncertainty, and
$-0.0177<d_{V}^{g}<0.0164$ and $-0.050<d_{A}^{g}<0.050$ assuming a 10$\%$ total uncertainty. It should be mentioned that the reported
bounds of each coupling are obtained when the other one is set to zero.

\section{Kinematic handle }\label{htsec}

In this section, we explore the
sensitivity of the $t\bar{t}\gamma\gamma$
process to probe the top-quark CMDM and CEDM by
looking into the kinematic distribution of final-state
particles. Equation~\ref{e1} indicates that additional terms originating from
dimension-six operators have a different
Lorentz structure as well as particular dependences on the field's
momentum. Thus, one expects that the rate and kinematic distribution of
final-state particles will be altered due to the presence of such anomalous
couplings. 

\begin{figure}[htb]
\begin{center}
\vspace{1cm}
\resizebox{0.4\textwidth}{!}{\includegraphics{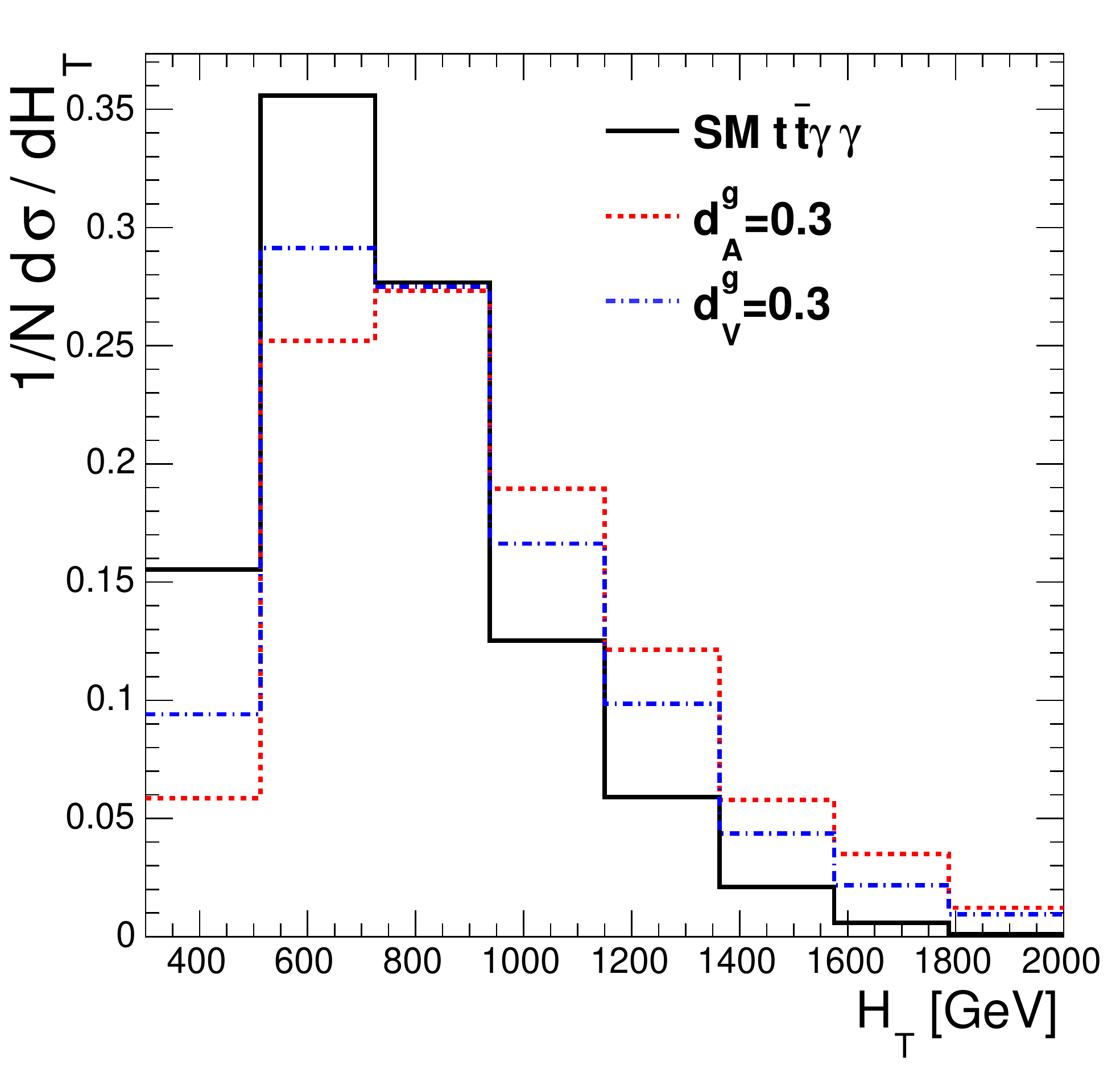}}  
\resizebox{0.4\textwidth}{!}{\includegraphics{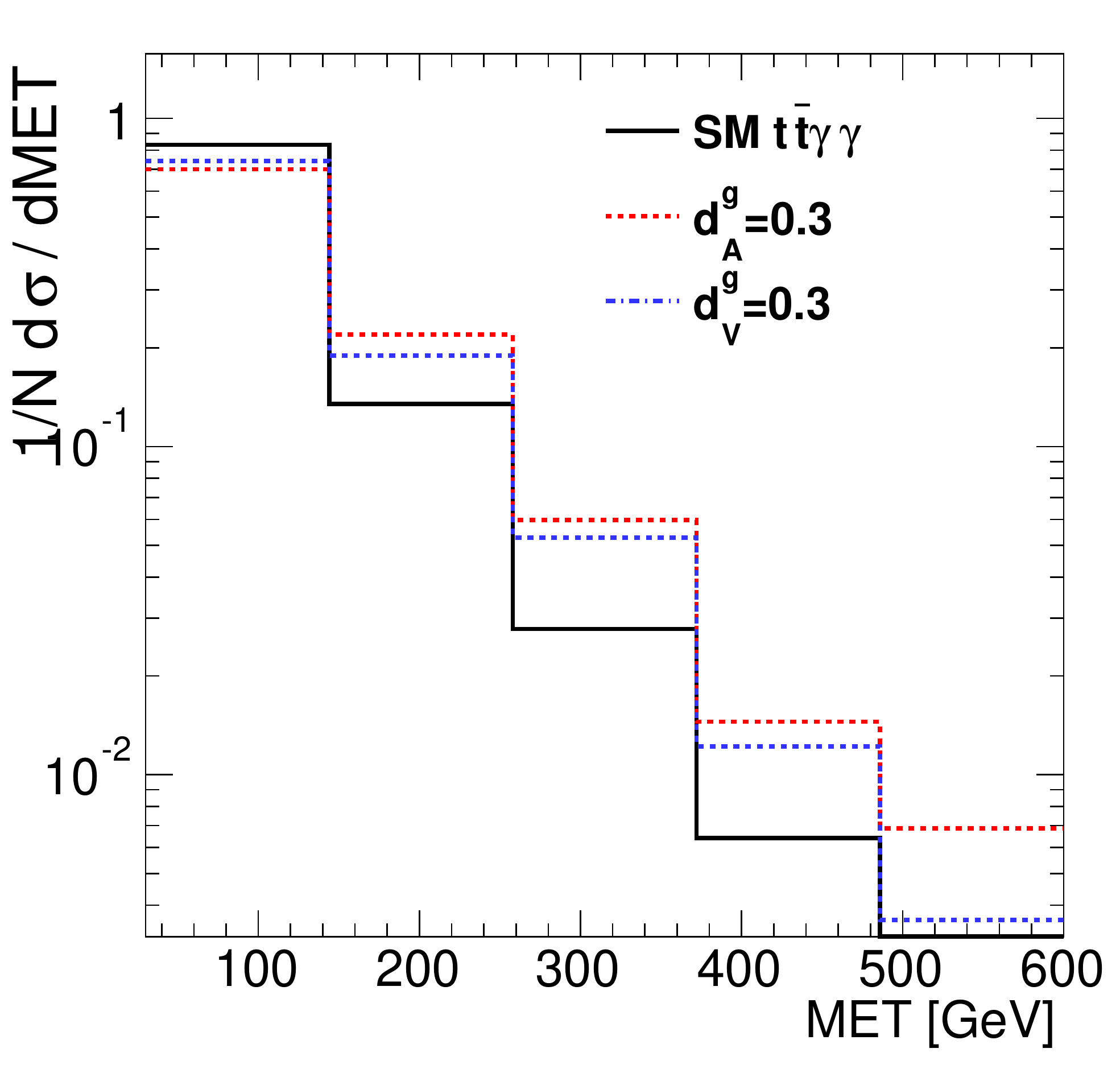}}  
\resizebox{0.4\textwidth}{!}{\includegraphics{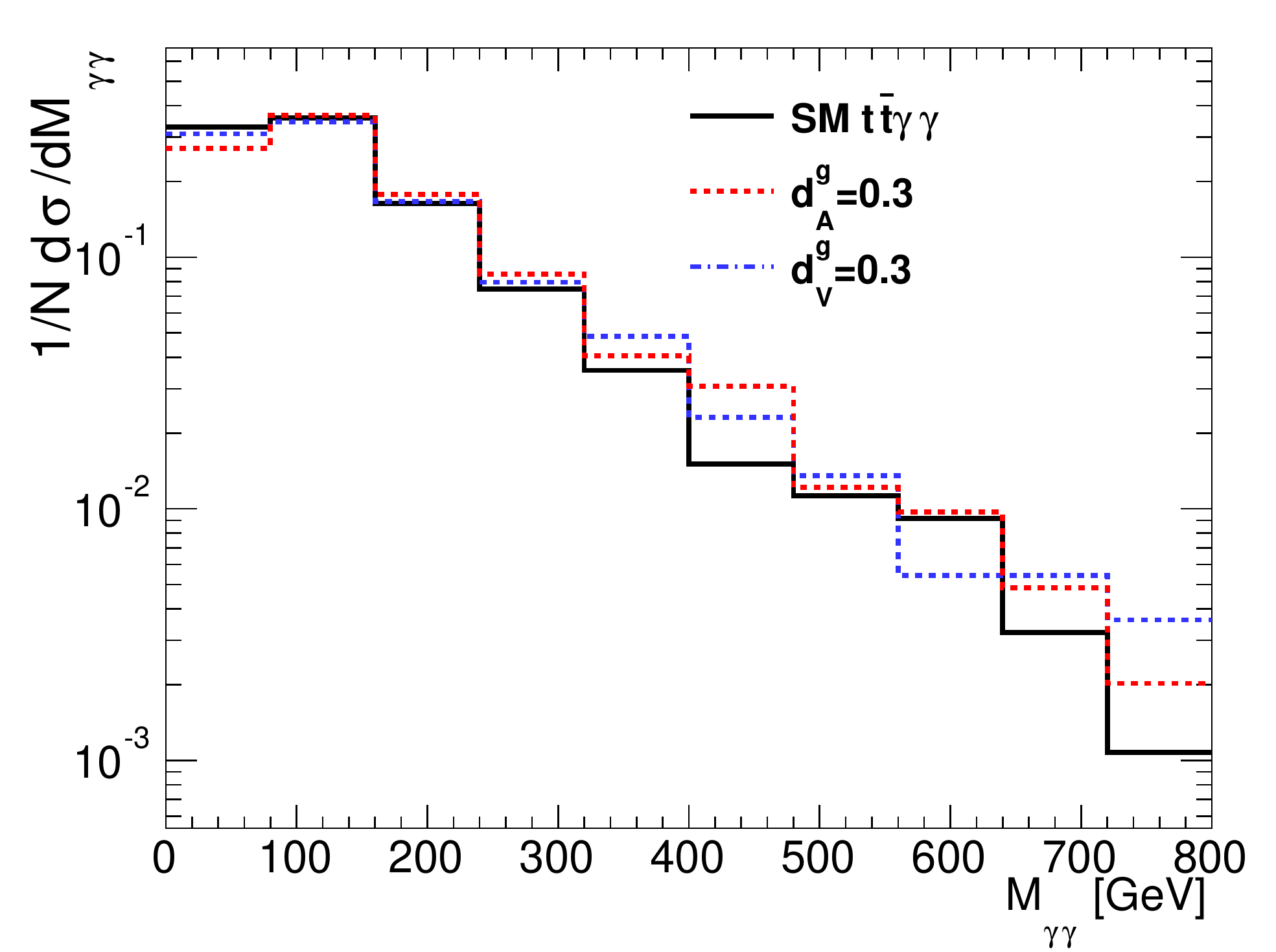}}  

\caption{Normalized distributions of $H_{T}$, missing transverse energy,
  and invariant mass of two photons. The plots compare the SM
  $t\bar{t}\gamma\gamma$ process with the same process when only
  either $d_{V}^{g}$ or  $d_{A}^{g}$ is applied.}\label{dists}
\end{center}
\end{figure}

In Figure~\ref{dists} we show some normalized kinematic distributions
to compare the expected SM $t\bar{t}\gamma\gamma$ process with the
same process when we apply only either $d_{V}^{g}$=0.3 or
$d_{A}^{g}$=0.3 each time.  The top left shows the $H_{T}$
distribution and the top right and bottom plots show the missing transverse energy and
invariant mass of two photons, respectively. Figure~\ref{dists}
indicates that including the new terms in the effective Lagrangian
modifies the shape of these distributions especially in the tail of
distributions where the process happens at higher energy
scale and shows the momentum dependence of these anomalous
couplings. It should be mentioned that these plots include the effects
of showering, hadronization, and object clustering, and detector
effects. 

We use the $H_{T}$ distribution as a sensitive
observable to find the potential  upper limit on the cross section of
$t\bar{t}\gamma\gamma$ in the presence of $d_{V}^{g}$ and $d_{A}^{g}$, and
then use this upper limit to obtain the constraint on these
couplings assuming that no deviation from the SM is observed. 
We use a single-bin counting experiment over the $H_{T}$ distribution in the signal region which
is the region with a high value of $H_{T}$. Essentially, this signal
region has to be optimized for the best cut value of $H_{T}$. The
conventional criteria are to obtain the value which results in the lowest limit on
the cross section or, in the other words the best power that bounds the
$d_{V}^{g}$ and $d_{A}^{g}$. Therefore, one needs to minimize the
95$\%$ expected limit on the signal cross section in order to find the
optimized $H_{T}$ value. It is worth mentioning that in the
optimization procedure one needs only to consider the statistical
uncertainty, and no systematic uncertainty is applied. The statistical
procedure to extract the expected limit is as follows. The probability of measuring N events in the signal region is
given by a Poisson distribution,

\begin{equation}
P(N|\,\sigma_{sig}\,\varepsilon\,\mathcal{L} ,
B)=e^{-(B+\sigma_{sig}\varepsilon\mathcal{L} )}\frac{(B+\sigma_{sig}\varepsilon\mathcal{L})^{N}}{N!},\label{eq:likelohood}
\end{equation}

where $\sigma_{sig}$, $\mathcal{L}$, $\varepsilon$, and $B$ are the signal
cross section, integrated luminosity, signal efficiency, and number of
expected background
events, respectively. These parameters are known except for the signal cross section, which is the
parameter of interest. The signal efficiency in the signal region is defined as the number of
events passing our selection cuts (explained in the section~\ref{analysis}) and
a certain cut value of $H_{T}$ over the total number of events that
only pass the $H_{T}$ cut. Exploiting the Bayesian approach, one can
extract the 95$\%$ C.L. upper limit on the signal cross
section in the signal region  by integrating over the posterior probability, defined as

\begin{equation}
0.95=\frac{\int_{0}^{\sigma^{95\%}}P(N|\,\sigma_{sig}\,\varepsilon\,\mathcal{L},
  B)d\sigma_{sig}}{\int_{0}^{\infty}P(N|\,\sigma_{sig}\,\varepsilon\,\mathcal{L},  B) d\sigma_{sig}}.\label{posterior}
\end{equation}

\begin{figure}[htb]
\begin{center}
\vspace{1cm}
\resizebox{0.45\textwidth}{!}{\includegraphics{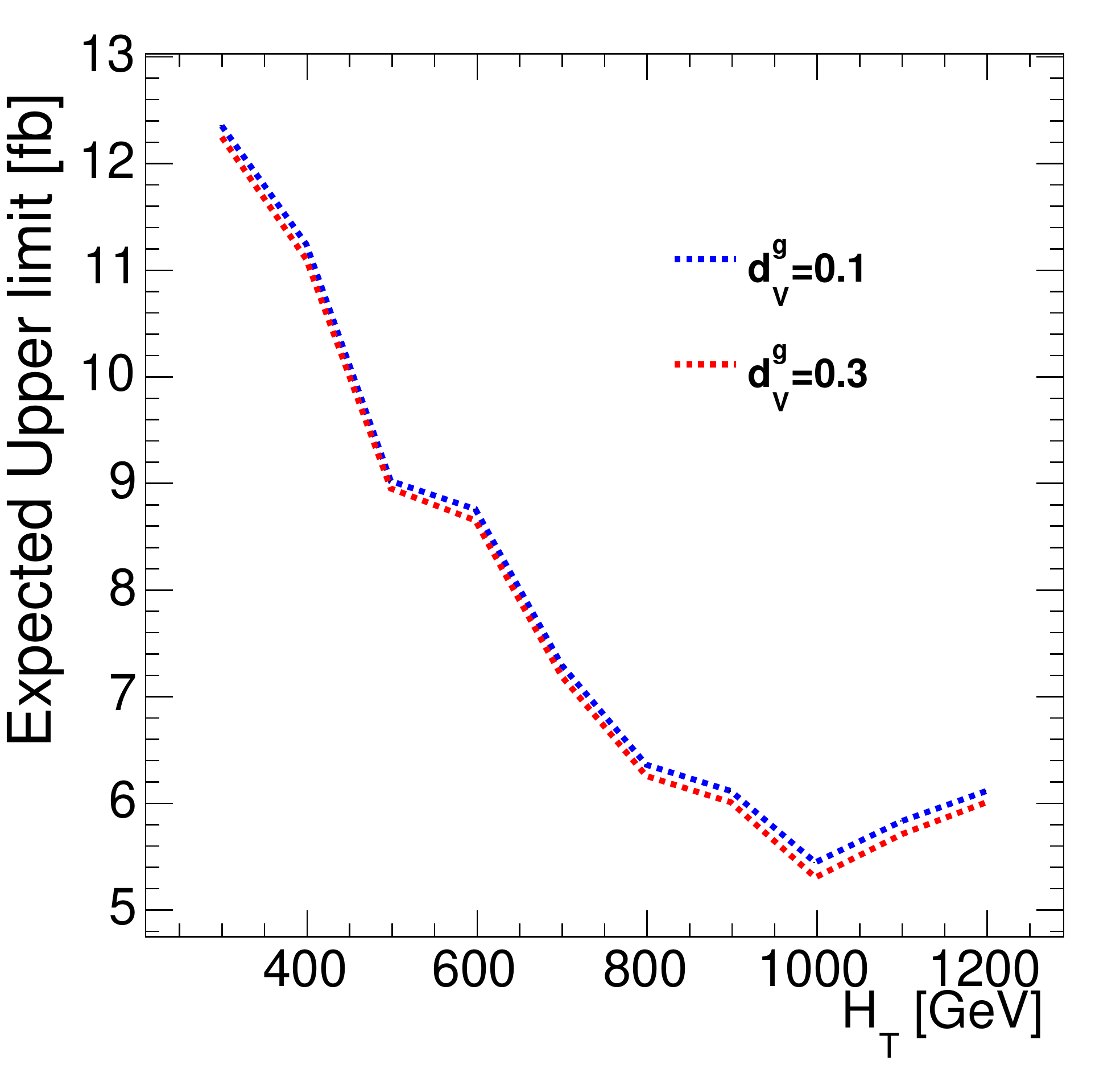}}  
\caption{ The 95$\%$ expected limit on the cross section as a function
of $H_{T}$ for two different values of $d_{V}^{g}$ is shown. The
optimized $H_{T}$ value is found to be 1000 GeV considering 100
$fb^{-1}$ integrated luminosity. }\label{optim}
\end{center}
\end{figure}

This statistical tool is employed to find the optimized cut
value for $H_{T}$. Therefore, we calculate the 95$\%$ C.L. expected limits on the
cross section for different values of $H_{T}$, ranging from 400 to
1200 GeV in steps of 100 GeV. The optimization is done separately when one
of the couplings is considered while the other coupling is set to
zero. This procedure is also performed for different values of each
coupling to see if any dependence on the coupling parameter exists. Figure~\ref{optim} shows the 95$\%$ expected limit as a
function of  $H_{T}$ for $d_{V}^{g}$ = 0.1, 0.3 considering 100
$fb^{-1}$ integrated luminosity.

The minimum expected
limit is obtained for $H_{T}$=1000 GeV. The same procedure is
implemented for $d_{A}^{g}$ = 0.1, 0.3 and the same optimized value is
obtained. Therefore, we consider the signal region by applying the
selection cuts and $H_{T}>$ 1000 GeV assuming 100 $fb^{-1}$ of data.

In the limit calculation procedure we consider the statistical and
systematic uncertainties due to the SM background processes. Given that,
most of these backgrounds have not been measured and we generate them at 
leading order, we assume $100\%$ uncertainty on the background yields in
the signal region.

We find the limits on $d_{V}^{g}$ and $d_{A}^{g}$ by comparing
the expected limit on the cross section with the theoretical cross
section in the signal region considering 100 and 300 $fb^{-1}$ at
$\sqrt{s}$= 13 TeV and 3000 $fb^{-1}$ at $\sqrt{s}$= 14 TeV.
 It should be mentioned that the compared theoretical cross section
curves in the signal region are subtracted from the SM value to
consider the pure non-SM cross sections originated by dimension-six operators.
The results obtained for the different integrated luminosities and different center-of-mass energies are shown in 
table~\ref{limitres}.  

\begin{table}[]
\centering
\caption{Obtained limit on $d_{V}^{g}$ and $d_{A}^{g}$
  considering 100, 300, and 3000 $fb^{-1}$ integrated luminosity }
\label{limitres}
\begin{tabular}{lll} \hline\hline
Integrated luminosity & $d_{V}^{g}$& $d_{A}^{g}$ \\ \hline\hline
100 $fb^{-1}$  (13 TeV) &        [ $-0.14,0.19$]       &     [ $-0.18,0.18$]        \\ \hline
300 $fb^{-1}$  (13 TeV)  &       [$ -0.10,0.15$]       &     [ $-0.13,0.13$]     \\ \hline
3000 $fb^{-1}$(14 TeV)&      [ $-0.006,0.03$]      &     [ $-0.014,0.014$]   \\   \hline \hline
\end{tabular}
\end{table}

\begin{figure}[htb]
\begin{center}
\vspace{1cm}
\resizebox{0.45\textwidth}{!}{\includegraphics{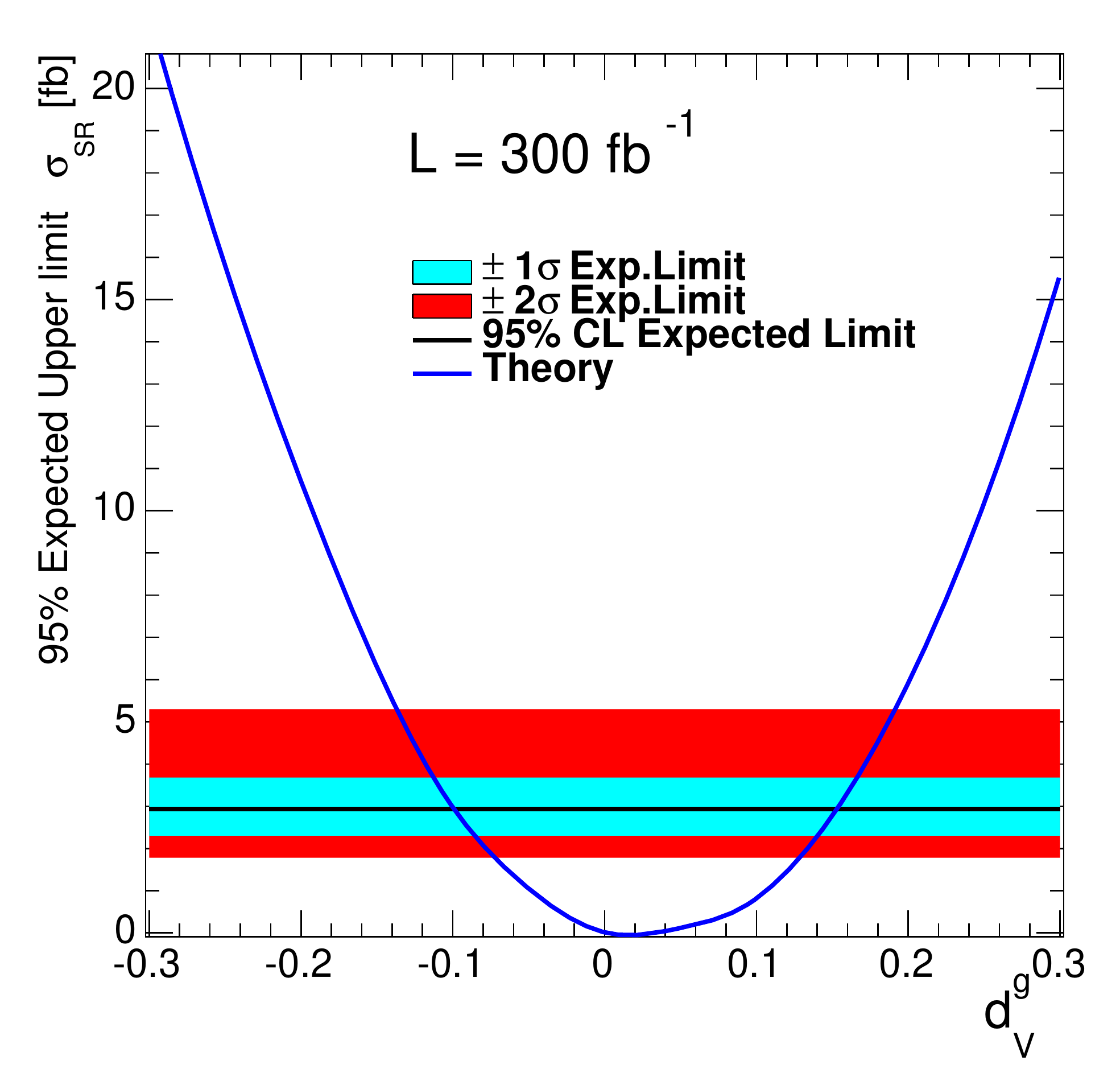}}  
\resizebox{0.45\textwidth}{!}{\includegraphics{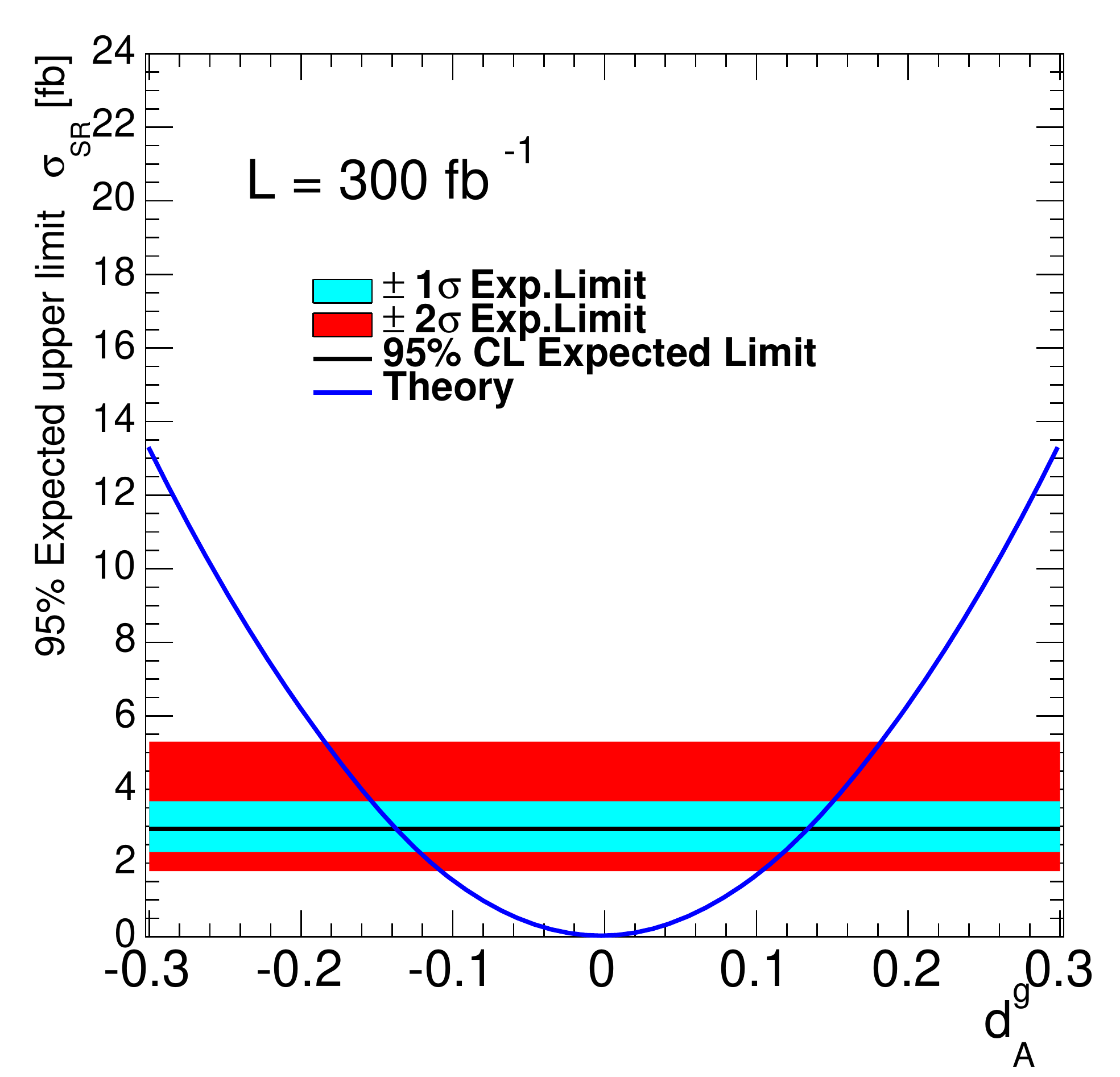}}  

\caption{ 95$\%$ C.L. expected upper limits on the signal cross section in
  the signal-dominated region compared with the theoretical cross
  sections of the signal for 300 $fb^{-1}$ integrated luminosity. }\label{limit}
\end{center}
\end{figure}

Figure~\ref{limit} shows the upper limits on 
$d_{V}^{g}$ (left) and $d_{A}^{g}$ (right) considering  300 $fb^{-1}$ integrated
luminosity which are compared with the theoretical curves.
The obtained bounds from the $H_{T}$ distribution
show very good  improvement using the 3000 $fb^{-1}$  expected amount
of data, especially for on the $d_{A}^{g}$ coupling.
 It should be mentioned that in general the
optimized cut value changes when one assumes different integrated
luminosities. Thus, in order to obtain the limit for each considered
amount of data, the optimization procedure is performed
separately. 

\begin{figure}[htb]
\begin{center}
\vspace{1cm}
\resizebox{0.47\textwidth}{!}{\includegraphics{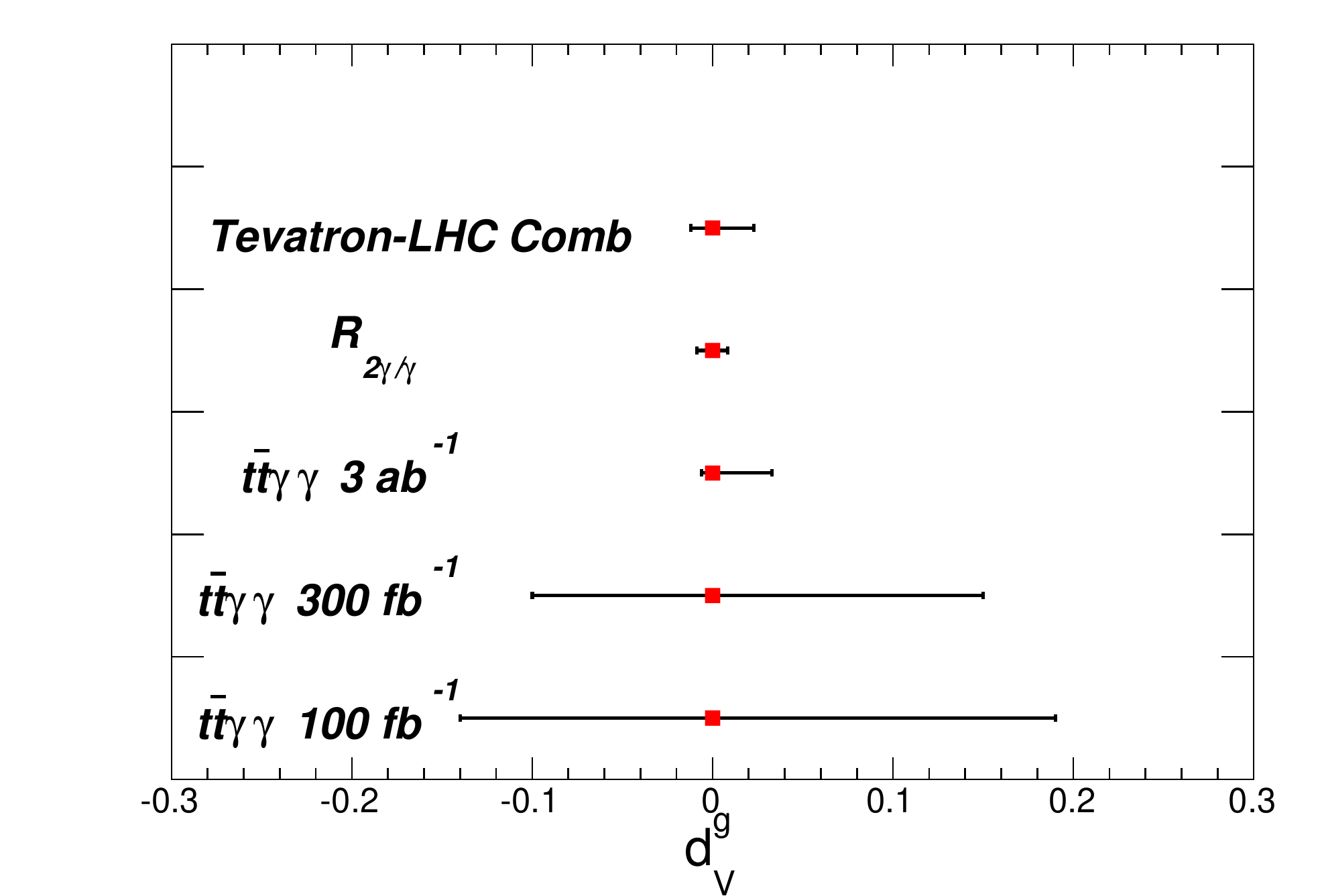}}  
\resizebox{0.47\textwidth}{!}{\includegraphics{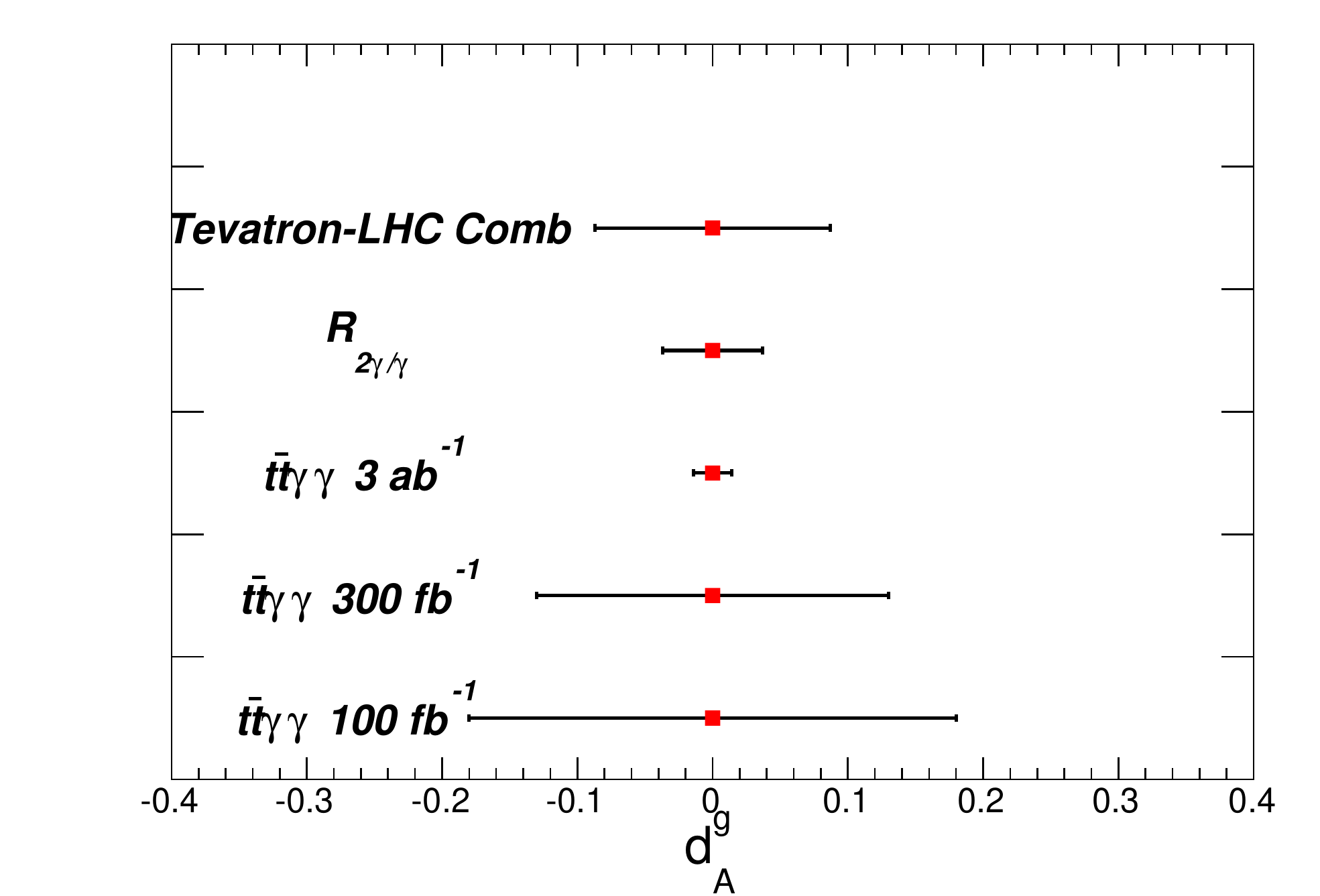}}  

\caption{ Summary of the
limits for  $d_{V}^{g}$ (left) and $d_{A}^{g}$ (right) obtained with
the different observables introduced in this
analysis assuming different integrated luminosities and the
combined results from the Tevatron and LHC8 are illustrated.}\label{summary}
\end{center}
\end{figure}

%
%
\section{Summary}\label{summary}

For the first time, rare SM top quark pair in association with two photons at
the LHC has been considered to investigate the prospects of constraining the top quark
chromo-moments. In the SM, these dipole moments are produced through the higher QCD,
electroweak loop corrections, which results in tiny values, and any
deviation from SM values would be a hint for new physics. In
addition, using the processes with high particle multiplicity helps to
effectively reduce the number of backgrounds. The
analysis was performed based on the effective  Lagrangian approach
where the dimension-six operators induced modifications to the $gt\bar{t}$
coupling. We considered the semi-leptonic decay of top quark pair and
defined a set of selection cuts to reconstruct this final state.

Then, a new cross section ratio in the selected phase space,
$R_{2\gamma/\gamma}=\sigma_{t\bar{t}\gamma\gamma}/\sigma_{t\bar{t}\gamma}$,
was introduced. This ratio is important in dealing with the top-quark
couplings for two reasons. First, in this observable, a considerable amount of theoretical and
experimental uncertainties cancel out. In addition to the
conventional reduction of uncertainties, the one
related to  photon identifications could be reduced due to the presence
of a photon in the both numerator and denominator. Second, due to the different
contributions of the gluon-gluon production mode in the
${t\bar{t}\gamma\gamma}$ and ${t\bar{t}\gamma}$ processes, the ratio can probe the different phase spaces of top quark
  couplings and effectively constrain the CP-violating
  coupling $d_{A}^{g}$. Considering a $5\%$ precision on
  this ratio measurement, we obtained the limits $-0.0088<d_{V}^{g}<0.0083$ and
$-0.037<d_{A}^{g}<0.037$. 

We also explored different kinematic distributions of final-state
particles, which include the effects of parton showering,
hadronization, jet clustering, and detector simulation. We selected
the distribution of the scalar sum of jet transverse energy, $H_{T}$. The contribution of
these non-SM couplings in the higher value of $H_{T}$ is pronounced
with respect to the pure SM contribution due to the dependence of the new
couplings on the momentum. We have optimized the $H_{T}$ cut value in
order to define a signal region where the best power to probe
these couplings is obtained. Finally, we used a counting bin
experiment method based on the Bayesian approach to find the upper limit
on the signal cross section in the signal region. By comparing the theoretical cross
section and upper limit in the defined signal region, we extracted
the bounds $-0.006<d_{V}^{g}<0.03$ and $-0.014 <d_{A}^{g}<0.014$ using
3 $ab^{-1}$ integrated luminosity. Figure~\ref{summary} shows the summary of the
limits for $d_{V}^{g}$ (left) and $d_{A}^{g}$ (right) obtained with
the different observables introduced in this
analysis, assuming different integrated luminosities and the combined results from Tevatron and LHC8.

\vspace{0.5cm}
%
{\bf Acknowledgments:}
We thank M. Mohammadi and S. Khatibi for their fruitful discussion,
comments, and help in implementing the model.
%

%
%

\end{document}